\documentclass[prb,twocolumn,showpacs,amsmath,amssymb]{revtex4}

\usepackage{graphicx}
\usepackage{dcolumn}% Align table columns on decimal point
\usepackage{bm}% bold math

\begin{document}

%% \preprint{APS/123-QED}

\title{
Mixed-state aspects of an out-of-equilibrium Kondo problem in a quantum dot 
}

\author{Akira Oguri}%
%\email{oguri@sci.osaka-cu.ac.jp}
\affiliation{Department of Material Science, Osaka City University, 
Sumiyoshi-ku, Osaka 558-8585, Japan}

\date{\today}

\begin{abstract}
We reexamine basic aspects of a nonequilibrium steady state 
in the Kondo problem for a quantum dot 
under a bias voltage ($eV = \mu_L-\mu_R$)  
 using a reduced density matrix (RDM). It is obtained in the Fock space 
 by integrating out one of the two conduction channels, 
which can be separated from the dot.
The remaining subspace is described by a single-channel Anderson model,
and the statistical distribution is determined by the RDM. 
At zero temperature, the noninteracting RDM is constructed as 
the mixed states that show a close similarity to 
the high-temperature distribution in equilibrium. 
Specifically, when the system has an inversion symmetry, 
the one-particle states in an energy region 
between the two chemical potentials ($\mu_R<\varepsilon<\mu_L$) 
are occupied, or unoccupied, completely at random with an equal weight. 
The Coulomb interaction preserves these aspects, 
and the correlation functions can be expressed 
in a Lehmann-representation form with the mixed-state distribution.
Furthermore, the statistical weight 
shows a broad distribution in the many-particle Hilbert space,
 spreading out in a wide energy region over 
 the bias energy  $eV$.

\end{abstract}


%\keywords{Suggested keywords}%Use showkeys class option if keyword
                              %display desired

\maketitle

\section{Introduction}
\label{sec:introduction}

The Kondo effect in quantum dots has been 
an active research field over a decade.\cite{GR,NG,Goldharber,Cronenwett,Simmel,VanDerWiel}
So far, the equilibrium and linear-response properties 
 of a single quantum dot have been understood well, basically, 
based on the knowledge of the Kondo physics 
in dilute magnetic alloys.\cite{Hewson} 
The nonlinear transport under a finite bias voltage $V$,\cite{WM,HDW} 
however, 
is still not fully understood, although a number of 
previous works\cite{TheoriesII,SchillerHershfield,KNG,KSL,OguriA,OguriB}
and recent ones\cite{FujiiUeda,GogolinKomnik,HB} 
have contributed to the continuous progress. 
One of the complications in the nonequilibrium system 
is that the density matrix is not a simple function 
of the Hamiltonian $H$. 
In thermal equilibrium, the density matrix has a universal form  
$\widehat{\rho}_{\rm eq}^{\phantom{0}} \propto e^{-\beta H}$. 
On the other hand, the nonequilibrium density matrix is not unique, 
even for the steady states.

The Keldysh formalism has widely been used for constructing  
the nonequilibrium density matrix.\cite{Keldysh,Caroli}
It has traditionally been described with a perturbative 
Green's function approach. However,
in order to study comprehensively the nonequilibrium 
properties of strongly correlated electron systems, 
accurate nonperturbative approaches are also required.
For instance, in the thermal equilibrium the approaches such 
as numerical renormalization group (NRG)\cite{KWW} 
and quantum Monte Carlo\cite{FH} methods played 
important roles to clarify the Kondo physics 
on the whole energy scale.
On the other hand, we still have not had systematic ways to 
handle the nonequilibrium density matrix. 
It makes the construction of nonperturbative approaches difficult. 
Therefore, further considerations about the basic formulation itself 
 seem to be still needed.

In this paper, we describe a reformulation of 
a nonequilibrium steady state, using a reduced density matrix (RDM).
It is obtained in the Fock space 
by integrating out one of the two channel degrees of freedom,
which can be separated out from the dot by a reconstruction 
of the linear combination of the channels.
The rest of the Hilbert space consists of 
the dot and the remaining channel, 
 which are coupled with each other by a tunneling matrix element.
The RDM takes a simple form in the noninteracting 
case, as shown in Eq.\ (\ref{eq:rho_s_0_T}). 
It enables us to treat the nonequilibrium averages  
using a single-channel Anderson model. 
The bias voltage appears in the subspace 
through an energy-dependent 
parameter $\mathcal{T}_{\varepsilon}$ given in Eq.\ (\ref{eq:T_eff}), 
which has a close relation to 
the usual nonequilibrium 
distribution function $f_\mathrm{eff}(\varepsilon)$. 
Specifically for the symmetric coupling $\Gamma_L=\Gamma_R$, 
the electrons distribute  at zero temperature 
to the one-particle states 
between the two chemical potentials $\mu_L$ and $\mu_R$ 
completely at random. 
The Coulomb interaction is taken into account adiabatically,
and then the nonequilibrium averages can be 
expressed in form of Eq.\ (\ref{eq:neq_average_u}),  
in terms of the interacting eigenstates and the RDM. 
This expression could be used for 
Hamiltonian-based nonperturbative calculations. 
It is also deduced that the nonequilibrium statistical weight 
 spreads out in a rather wide energy region of 
the many-particle Hilbert space over the bias energy $eV$.

The model is described in Sec.\ \ref{sec:formulation}.
The derivation of the RDM is given in Sec.\ \ref{sec:RDM}.
The mixed-state aspects of the RDM are examined 
for $\Gamma_L=\Gamma_R$ in Sec.\ \ref{sec:mixed-state}.
The nonequilibrium distribution in the many-particle Hilbert space 
is discussed in Sec.\ \ref{sec:average}.
Summary is given in Sec.\ \ref{sec:summary}.
In the Appendixes, details of some calculations are provided.

\section{Formulation}
\label{sec:formulation}

\subsection{Scattering states}

We start with a single Anderson impurity 
connected to two leads on the left ($L$) and right ($R$).
The Hamiltonian is given by  $H = H_0 + H_{U}$ with 
 $H_0 = H_{d} + H_{c} + H_T$, 
\begin{align}
 &H_{d}\,   = \,   
\Bigl( \epsilon_d + \frac{U}{2}\,  \Bigr)\,  n_{d}^{\phantom 0} \,, 
\qquad 
 H_{U}  =  \frac{U}{2} \Bigl(n_{d}^{\phantom 0} -1\Bigr)^2 ,
\label{eq:Hd}
\\
  & H_{c}  \,=  
\sum_{\nu=L,R}\sum_{\sigma} 
\int  \! d\varepsilon\,  \varepsilon\, 
c^{\dagger}_{\varepsilon,\nu\sigma} \,
c^{\phantom{\dagger}}_{\varepsilon,\nu\sigma} 
\;,
\label{eq:Hc}
\\
& H_T \,  =\,
\sum_{\nu=L,R}\sum_{\sigma} 
 v_{\nu}^{\phantom{0}} \, \Bigl( 
\,d_{\sigma}^{\dagger} \,C_{\nu\sigma}^{\phantom{\dagger}}
+
C_{\nu\sigma}^{\dagger}
\,d_{\sigma}^{\phantom{\dagger}} \Bigr)
\;.
\label{eq:Hmix}
\end{align}
Here,  
 $d_{\sigma}^{\dagger}$ creates an electron with spin 
$\sigma$ in the dot, 
$n_{d}^{\phantom 0} = 
\sum_{\sigma} d^{\dagger}_{\sigma} d^{\phantom{\dagger}}_{\sigma}$,
 $\epsilon_d$ is the onsite potential, and  
$v_{\nu}^{\phantom{0}}$ is 
a tunneling matrix element between the dot and lead $\nu$. 
The conduction electrons are normalized as  
$\{ c^{\phantom{\dagger}}_{\varepsilon,\nu\sigma}, 
c^{\dagger}_{\varepsilon',\nu'\sigma'}
\} = \delta_{\nu\nu'} \,\delta_{\sigma\sigma'}\, 
\delta(\varepsilon-\varepsilon')$, and 
 $C_{\nu\sigma}^{\phantom{\dagger}} = 
\int   d\varepsilon   
\sqrt{\rho(\varepsilon)} 
\, c^{\phantom{\dagger}}_{\varepsilon,\nu\sigma}$ 
with a condition $\int d\varepsilon\, \rho(\varepsilon) =1$;   
the same one-particle density of states
$\rho(\varepsilon)$ is assumed for both of the leads.
We are using explicitly the continuous form of the conduction bands\cite{KWW} 
in order to make the scattering states well defined. 
The Fermi level at equilibrium is taken to be $\mu = 0$.
The two different chemical potentials, $\mu_L=-\mu_R=eV/2$, 
are introduced usually into the isolated leads for $H_T=0$.\cite{Caroli} 
Alternatively, as described by Hershfield,\cite{Hershfield} 
the same steady state can be constructed 
from $H_0$ including $H_T$ in the initial condition, 
and it has been applied to 
some special models.\cite{SchillerHershfield,Han} 
We use the later formulation in the following 
to set up the nonequilibrium density matrix,
and assign $\mu_L$ and $\mu_R$ to 
the scattering states incident from left and right, 
respectively.
We will use units $\hbar=1$, except for 
the current defined in Eq.\ (\ref{eq:current_op}).

The two conduction bands can be classified into  
the $s$-wave and $p$-wave parts, defined by 
\begin{align} 
&s_{\varepsilon\sigma}^{\phantom{\dagger}} 
=
\frac{
v_R^{\phantom{|}} c_{\varepsilon,R\sigma}^{\phantom{\dagger}}
+v_L^{\phantom{|}} 
c_{\varepsilon,L\sigma}^{\phantom{\dagger}}}{\sqrt{v_R^2+v_L^2}} 
\,, \quad 
p_{\varepsilon\sigma}^{\phantom{\dagger}} 
=
\frac{
v_L^{\phantom{|}} c_{\varepsilon,R\sigma}^{\phantom{\dagger}}
-v_R^{\phantom{|}} 
c_{\varepsilon,L\sigma}^{\phantom{\dagger}}}{\sqrt{v_R^2+v_L^2}} 
\,.
\label{eq:s_p_def}
\end{align} 
The $s$-wave electrons couple to the dot 
via the tunneling matrix elements, as 
$H_T= 
\sum_{\sigma} 
\overline{v} \left( 
d_{\sigma}^{\dagger} S_{\nu\sigma}^{\phantom{\dagger}}
+
S_{\nu\sigma}^{\dagger}
d_{\sigma}^{\phantom{\dagger}} \right)
$, where
 $S_{\sigma}^{\phantom{\dagger}} = 
\int   d\varepsilon   
\sqrt{\rho(\varepsilon)} 
\, s^{\phantom{\dagger}}_{\varepsilon\sigma}$  
and $\overline{v} \equiv \sqrt{v_R^2+v_L^2}$.
The $p$-wave electrons do not couple to the dot, 
but they contribute to the current 
 through the impurity,\cite{KNG}  
\begin{align} 
\!\!
J  \equiv 
 \frac{v_R^2 J_L + v_L^2 J_R }{v_R^2+v_L^2} 
% \nonumber \\
=
%&\,
-i \frac{e}{\hbar} 
\frac{v_L^{\phantom{|}}v_R^{\phantom{|}}}
{\overline{v}}
\sum_{\sigma}
\left( 
P_{\sigma}^{\dagger}
d_{\sigma}^{\phantom{\dagger}} 
-
d_{\sigma}^{\dagger} 
P_{\sigma}^{\phantom{\dagger}}
\right) .
\label{eq:current_op}
\end{align}
Here, 
$P_{\sigma}^{\phantom{\dagger}} = 
\int   d\varepsilon   
\sqrt{\rho(\varepsilon)} 
\, p^{\phantom{\dagger}}_{\varepsilon\sigma}$, and 
 $J_L$ ($J_R$) is the current flowing
from the left lead (dot) to the dot (right lead). 
The scattering states in the $s$-wave part, 
$\alpha_{\varepsilon\sigma}^{\dagger}$, 
can be calculated explicitly for $H_U=0$ 
including all effects of $H_T$,  
\begin{align} 
&\alpha_{\varepsilon\sigma}^{\dagger} 
\,= \,
 s_{\varepsilon\sigma}^{\dagger}
\,+\,
\overline{v}\,
\sqrt{\rho(\varepsilon)}
\,G^{r}_{0}(\varepsilon) 
\,d_{\sigma}^{\dagger} \,
 \nonumber
 \\ 
&  \qquad \qquad \ \ + \overline{v} \sqrt{\rho(\varepsilon)}
\,G^{r}_{0}(\varepsilon)\!
\int \! d\varepsilon'\,
\frac{\sqrt{\rho(\varepsilon')}}{\varepsilon - \varepsilon' + i \delta} 
\ s_{\varepsilon'\sigma}^{\dagger}
\,, \\
& G^{r}_{0}(\varepsilon)
=\, \left[ \varepsilon- \left(\epsilon_d +\frac{U}{2}\right)
 -\overline{v}^2 \!\!\int\! d\varepsilon' 
\frac{\rho(\varepsilon')}{\varepsilon-\varepsilon' +i \delta} 
\right]^{-1} \! .
\label{eq:G^r_0}
\end{align} 
Note that 
the resonance width of the impurity level is given by  
 $\Delta(\varepsilon) 
\equiv  \pi \overline{v}^2  \rho(\varepsilon) =\Gamma_L+\Gamma_R$ 
with $\Gamma_{\nu} = \pi v_{\nu}^2  \rho(\varepsilon)$ 
in the noninteracting case.
We will also be using the advanced Green's function 
$G^{a}_{0}(\varepsilon)
 \equiv \left\{ G^{r}_{0}(\varepsilon) \right\}^*$. 
The one-particle states $\alpha^{\phantom{\dagger}}_{\varepsilon\sigma}$
and $p^{\phantom{\dagger}}_{\varepsilon\sigma}$ constitute
the eigenstates of $\,H_0 =H_0^S+H_0^P$,   
\begin{align}
H_0^S 
= 
\sum_{\sigma} \!
\int  \!\! d\varepsilon\,  \varepsilon\, 
\alpha^{\dagger}_{\varepsilon\sigma} 
\alpha^{\phantom{\dagger}}_{\varepsilon\sigma} 
, \quad \ 
H_0^P 
=
\sum_{\sigma}\! 
\int  \!\! d\varepsilon\,  \varepsilon\, 
p^{\dagger}_{\varepsilon\sigma} 
p^{\phantom{\dagger}}_{\varepsilon\sigma} 
\, .
\label{eq:H0_Jost1} 
\end{align} 
The scattering state is also normalized as  
$\{ \alpha^{\phantom{\dagger}}_{\varepsilon,\sigma}, 
\alpha^{\dagger}_{\varepsilon',\sigma'}
\} = \delta_{\sigma\sigma'}\, 
\delta(\varepsilon-\varepsilon')$, and 
the impurity state $d_{\sigma}$ can be expressed in terms of 
 $\alpha_{\varepsilon\sigma}$ as
\begin{align}
d_{\sigma} =
\overline{v} \int d\varepsilon \sqrt{\rho(\varepsilon)}\,  
G_{0}^{r}(\varepsilon) \,
\alpha^{\phantom{\dagger}}_{\varepsilon\sigma}\;.
\label{eq:d}
\end{align} 
Here, we have omitted, for simplicity,
the possibility that bound states could emerge 
outside of the conduction band.
It can be taken into account when it is necessary.

\subsection{Nonequilibrium steady state}

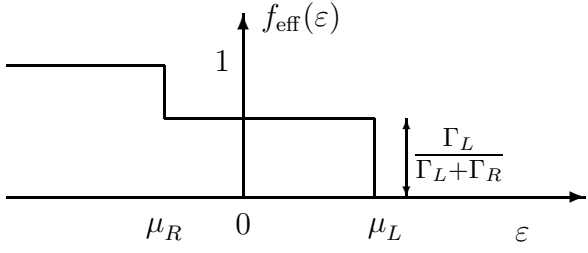
\begin{figure}[tb]
% \begin{center}
\setlength{\unitlength}{0.7mm}
%% \setlength{\unitlength}{0.6mm}
%% \begin{minipage}{1\linewidth}
 \hspace{-1.8cm}
\begin{picture}(105,49)(-56,-10)
\thicklines

\put(-45,0){\vector(1,0){110}}
\put(0,0){\vector(0,1){35}}

\put(-45,25){\line(1,0){30}}
\put(-15,15){\line(0,1){10}}
\put(-15,15){\line(1,0){40}}
\put(25,0){\line(0,1){15}}

\put(0,-5){\makebox(0,0){\large $0$}}
\put(-15,-5){\makebox(0,0){\large $\mu_R^{\phantom{0}}$}}
\put(27,-5){\makebox(0,0){\large $\mu_L^{\phantom{0}}$}}
\put(54,-7){\makebox(0,0){\large $\varepsilon$ }}
\put(11,34){\makebox(0,0){\large $f_{\mathrm{eff}}(\varepsilon)$}}
\put(-4,26){\makebox(0,0){\large $1$}}
\put(41,8){\makebox(0,0)
   {\Large $\frac{\mathstrut \Gamma_L}{\mathstrut \Gamma_L+\Gamma_R}$}}

\thinlines

\put(31,10){\vector(0,-1){10}}
\put(31,10){\vector(0,1){5}}

\end{picture}
%% \end{minipage}
\caption{
Distribution function $f_{\mathrm{eff}}(\varepsilon)$
at $T=0$.}
\label{fig:distribution}
% \end{center}
\end{figure}

The noninteracting Hamiltonian can also be diagonalized 
with the scattering states moving towards left and right,  
\begin{align} 
&H_0 = 
\sum_{\nu=L,R} \sum_{\sigma}
 \int  \! d\varepsilon\,  \varepsilon \,
 \gamma^{\dagger}_{\varepsilon,\nu\sigma} 
 \gamma^{\phantom{\dagger}}_{\varepsilon,\nu\sigma}
\;,
\label{eq:H0_Jost2} 
\\
&\gamma_{\varepsilon,R\sigma}^{\phantom{\dagger}} 
=
\frac{
v_R^{\phantom{|}}\,\alpha_{\varepsilon\sigma}^{\phantom{\dagger}} 
+ v_L^{\phantom{|}}\, 
p_{\varepsilon\sigma}^{\phantom{\dagger}}}{\sqrt{v_L^2+v_R^2}} 
\;, 
\quad 
\gamma_{\varepsilon,L\sigma}^{\phantom{\dagger}} 
=
\frac{
v_L^{\phantom{|}}\,
\alpha_{\varepsilon\sigma}^{\phantom{\dagger}} 
- v_R^{\phantom{|}}\, 
p_{\varepsilon\sigma}^{\phantom{\dagger}}}{\sqrt{v_L^2+v_R^2}} 
\;.
\label{eq:scattering_state}
\end{align}
This propagating-wave form of $H_0$ 
are compatible with 
the standing-wave form Eq.\ (\ref{eq:H0_Jost1}),  
as the scattering states have two-fold degeneracy. 
The chemical potentials $\mu_L$ and $\mu_R$ can 
be assigned to these two propagating states 
$\gamma^{\dagger}_{\varepsilon,L\sigma}$ and 
$\gamma^{\dagger}_{\varepsilon,R\sigma}$, respectively,
using\cite{Hershfield,KNG} 
\begin{align}
Y_0 \equiv&\,
\frac{eV}{2}
\sum_{\sigma}
\int\!d\varepsilon 
\left(
\gamma^{\dagger}_{\varepsilon,L\sigma} 
\gamma^{\phantom{\dagger}}_{\varepsilon,L\sigma} 
-
\gamma^{\dagger}_{\varepsilon,R\sigma} 
\gamma^{\phantom{\dagger}}_{\varepsilon,R\sigma} 
\right)
\nonumber 
\\
=&\,
-\, 
eV \, \frac{v_Lv_R}{v_L^2+v_R^2}
\sum_{\sigma}
\int\!d\varepsilon \,
\Bigl( 
\alpha^{\dagger}_{\varepsilon\sigma} 
\,p^{\phantom{\dagger}}_{\varepsilon\sigma} 
+
p^{\dagger}_{\varepsilon\sigma} 
\alpha^{\phantom{\dagger}}_{\varepsilon\sigma} 
\Bigr) 
\nonumber
\\
&\, + 
\frac{eV}{2}  \frac{v_L^2-v_R^2}{v_L^2+v_R^2}
\sum_{\sigma}
\int\!d\varepsilon \,
\Bigl( 
\alpha^{\dagger}_{\varepsilon\sigma} 
\alpha^{\phantom{\dagger}}_{\varepsilon\sigma} 
-
p^{\dagger}_{\varepsilon\sigma} 
p^{\phantom{\dagger}}_{\varepsilon\sigma} 
\Bigr) .
\label{eq:Y_0}
\end{align}
With this operator, the noninteracting density matrix can be constructed as
\begin{align}
 \widehat{\rho}_0^{\phantom{0}} = 
\frac{ e^{ -\beta (H_0- Y_0 )}}{\Xi_0} \;,
\qquad \Xi_0 \equiv \mbox{Tr}\, 
 e^{ -\beta (H_0- Y_0 )}\;.
\label{eq:rho_0}
\end{align}
The statistical average is defined by  
$\langle \cdots\rangle_0 = 
\mbox{Tr}\,[\widehat{\rho}_0^{\phantom{0}} \cdots]$. 
For instance, the distribution function 
for $\gamma^{\phantom{\dagger}}_{\varepsilon,\nu\sigma}$ is given by
\begin{align} 
\langle 
\gamma^{\dagger}_{\varepsilon,\nu\sigma} 
\gamma^{\phantom{\dagger}}_{\varepsilon',\nu'\sigma'} 
\rangle_0 = f_{\nu}(\varepsilon )\,
\delta(\varepsilon-\varepsilon')\,
\delta_{\nu\nu'}\,
\delta_{\sigma\sigma'}\;,
\label{eq:gamma_average}
\end{align} 
where $f_{\nu}(\varepsilon) \equiv f(\varepsilon- \mu_{\nu})$, 
and $f(\varepsilon)=\left[ e^{\varepsilon/T}+1 \right]^{-1}$ is 
the Fermi function.
Similarly, 
from Eqs.\ (\ref{eq:scattering_state}) and (\ref{eq:gamma_average}), 
we obtain  
\begin{align} 
 \langle 
\alpha^{\dagger}_{\varepsilon\sigma} 
\alpha^{\phantom{\dagger}}_{\varepsilon'\sigma'} 
\rangle_0 
\,=& 
\,
f_\mathrm{eff}(\varepsilon)\,
\delta(\varepsilon-\varepsilon')\,
\delta_{\sigma\sigma'}\;,
\label{eq:f_eff}
\\
f_\mathrm{eff}(\varepsilon) \equiv& \,
\frac{\Gamma_L\,f_L(\varepsilon) + \Gamma_R\,f_R(\varepsilon)}
{\Gamma_L + \Gamma_R} \;,  
\label{eq:f_eff_def}
 \end{align} 
 and
 \begin{align} 
& \langle 
 \alpha^{\dagger}_{\varepsilon\sigma} 
p^{\phantom{\dagger}}_{\varepsilon'\sigma'} 
\rangle_0 
\,=\, \langle 
p^{\dagger}_{\varepsilon'\sigma'} 
 \alpha^{\phantom{\dagger}}_{\varepsilon\sigma} 
\rangle_0 
\nonumber\\
& \quad 
= 
-\,
\frac{v_L^{\phantom{|}}v_R^{\phantom{|}}}
{v_R^2+v_L^2}
\bigl[\,
f_L(\varepsilon) -
f_R(\varepsilon)\,\bigr]\,
\delta(\varepsilon-\varepsilon')\,
\delta_{\sigma\sigma'} .
\label{eq:alpha_p}
\end{align} 
Here, the distribution function $f_\mathrm{eff}(\varepsilon)$
appears not only for the impurity state
but also for all the scattering states with  
the $s$-wave character.\cite{HDW}
At zero temperature,  $f_\mathrm{eff}(\varepsilon)$ 
takes a constant value $\Gamma_L/(\Gamma_R+\Gamma_L)$ 
 for $\mu_R<\varepsilon<\mu_L$, 
as shown in Fig.\ \ref{fig:distribution}. 
Specifically, the constant becomes $1/2$ for 
the symmetric coupling $\Gamma_L=\Gamma_R$. 
This feature looks quite similar to 
a high-temperature behavior of the Fermi function  
$\left. f(\varepsilon)\right|_{T\to\infty} =1/2$, 
and it can be explained  
as a manifestation of the mixed-state nature of 
the RDM, which is given in Eq.\ (\ref{eq:rho_s_0_T}).

%------------------------------------------------------------

%--------------------------(fig1_feff.tex)-----------------------
% \begin{figure}[tb]
% \setlength{\unitlength}{0.7mm}
%  \hspace{-1.8cm}
% \begin{picture}(105,49)(-56,-10)
% \thicklines
% 
% \put(-45,0){\vector(1,0){110}}
% \put(0,0){\vector(0,1){35}}
% 
% \put(-45,25){\line(1,0){30}}
% \put(-15,15){\line(0,1){10}}
% \put(-15,15){\line(1,0){40}}
% \put(25,0){\line(0,1){15}}
% 
% \put(0,-5){\makebox(0,0){\large $0$}}
% \put(-15,-5){\makebox(0,0){\large $\mu_R^{\phantom{0}}$}}
% \put(27,-5){\makebox(0,0){\large $\mu_L^{\phantom{0}}$}}
% \put(54,-7){\makebox(0,0){\large $\varepsilon$ }}
% \put(11,34){\makebox(0,0){\large $f_{\mathrm{eff}}(\varepsilon)$}}
% \put(-4,26){\makebox(0,0){\large $1$}}
% \put(41,8){\makebox(0,0)
%    {\Large $\frac{\mathstrut \Gamma_L}{\mathstrut \Gamma_L+\Gamma_R}$}}
% 
% \thinlines
% 
% \put(31,10){\vector(0,-1){10}}
% \put(31,10){\vector(0,1){5}}
% 
% \end{picture}
% \caption{
% Distribution function $f_{\mathrm{eff}}(\varepsilon)$
% at $T=0$.}
% \label{fig:distribution}
% \end{figure}
%------------------------------------------------------------

 The Coulomb interaction is  switched on adiabatically 
 based on the Keldysh formalism,\cite{Keldysh} 
\begin{align} 
&\widehat{\rho}
 \equiv S(0,-\infty) \,\widehat{\rho}_0^{\phantom{0}}\,
  S(-\infty,0) \;,
\label{eq:rho_org}
\\ 
&S(t_2,t_1) \equiv \mbox{T} \exp \! 
\left[- i \! \int_{t_1}^{t_2} \! dt\, 
e^{iH_0^S t} H_U e^{-iH_0^S t} \right].
\label{eq:S}
\end{align} 
Note that 
 $\bigl[ S(0,-\infty), \,
p_{\varepsilon\sigma}^{\phantom{\dagger}} \bigr]=0$,
 and thus the interaction affects only the $s$-wave part 
of the Hilbert space, which is described by the Hamiltonian
\begin{equation}
H^S= H_0^S+H_U \;.
\label{eq:H_s}
\end{equation}
Hershfield has provided one possible way 
to construct an interacting operator $Y$, 
by which the full density matrix is written  
in the form $\widehat{\rho} \propto e^{-\beta(H-Y)}$.\cite{Hershfield}
In the present study, however,
we go back to the original definition of $\widehat{\rho}$ given 
in Eq.\ (\ref{eq:rho_org}) for a general description.

The interacting average $\langle \cdots\rangle$ is taken with 
the full density matrix $\widehat{\rho}$.
For instance, 
using Eq.\ (\ref{eq:d}),
the average for the local charge can be written in the form
\begin{align} 
\langle 
 n_d^{\phantom{0}} 
\rangle
=& \ 
\overline{v}^2 
\sum_{\sigma}
\!\int \!\!d\varepsilon\,d\varepsilon' 
\sqrt{\rho(\varepsilon')\rho(\varepsilon)} \,
G^{0}_{a}(\varepsilon') G^{0}_{r}(\varepsilon)
\bigl\langle 
\alpha_{\varepsilon' \sigma}^{\dagger} 
\alpha_{\varepsilon \sigma}^{\phantom{\dagger}} 
\bigr\rangle ,
\nonumber \\
=&
\sum_{\sigma}
\!\int \! \frac{d\omega}{2 \pi i} \,  G^{<}(\omega) \;. 
\label{eq:nd_GF}
 \end{align}
Here, $ 
G^{<}(\omega) 
\equiv i \int_{-\infty}^{\infty} \! dt \, e^{i \omega t}
\bigl\langle 
d_{\sigma}^{\dagger}\,
e^{iH^S t}d_{\sigma}^{\phantom{\dagger}}e^{-iH^S t}  
 \bigr\rangle$ 
is the lesser Green's function,
which for the noninteracting limit is given by 
$
G_{0}^{<}(\varepsilon) =  
-f_{\mathrm{eff}}(\varepsilon)
        \left[ G_{0}^r(\varepsilon) 
- G_{0}^a(\varepsilon) \right]$.
The nonequilibrium current can also be expressed 
as an average defined in the $s$-wave subspace,\cite{MW} 
\begin{align}
\langle J \rangle =&
 -i\, \frac{e\,v_L^{\phantom{|}} v_R^{\phantom{|}}}{\hbar}
\sum_{\sigma}
\int\!\!
d\varepsilon \,
d\varepsilon' 
\sqrt{\rho(\varepsilon') \rho(\varepsilon)} 
\nonumber \\
&  \times 
\left[\,
G_{0}^{r}(\varepsilon)\,
\bigl\langle p_{\varepsilon'\sigma}^{\dagger} 
\alpha_{\varepsilon\sigma}^{\phantom{\dagger}}  \bigr\rangle 
-
G_{0}^{a}(\varepsilon)\,
\bigl\langle \alpha_{\varepsilon\sigma}^{\dagger} 
p_{\varepsilon'\sigma}^{\phantom{\dagger}}  \bigr\rangle 
\,\right] 
% \label{eq:current_GF}
\;, 
\nonumber \\
=&\,  \frac{2 e}{h}  \int \!\! d\omega 
  \,  \bigl[ f_L(\omega) - f_R(\omega) \bigr] 
\frac{4
\Gamma_L
\Gamma_R
}{\Gamma_L
+\Gamma_R
}
 \left[\, - 
 {\rm Im}\, G^r(\omega) \,\right] .
\label{eq:caroli}
\end{align}
Here,
$ 
G^r(\omega) 
\equiv -i \int_0^{\infty} \! dt \, e^{i(\omega+i\delta)t}
\bigl\langle 
\bigl\{e^{iH^S t}d_{\sigma}^{\phantom{\dagger}}e^{-iH^S t},  
d_{\sigma}^{\dagger} \bigr\}  \bigr\rangle$ 
is the full retarded Green's function. 
Therefore, to calculate these averages,  
the $p$-wave degrees of freedom can be integrated out 
first before taking the trace over the correlated $s$-wave part.

\section{Reduced Density Matrix}
\label{sec:RDM}

We describe in this section a derivation of the RDM.
To this end, we introduce a discretized Hamiltonian 
corresponding to Eq.\ (\ref{eq:H0_Jost1})
as 
$\widetilde{H}_0 = \widetilde{H}_0^S + \widetilde{H}_0^P$,  
\begin{align}
&
\widetilde{H}_0^S  
 = \! \sum_{\sigma}  \sum_{m}   \varepsilon_m \,
\Bigl( 
\widetilde{\alpha}^{\dagger}_{\varepsilon_m,\sigma} 
 \widetilde{\alpha}^{\phantom{\dagger}}_{\varepsilon_m,\sigma}
-
\langle
\widetilde{\alpha}^{\dagger}_{\varepsilon_m,\sigma} 
 \widetilde{\alpha}^{\phantom{\dagger}}_{\varepsilon_m,\sigma}
\rangle^\mathrm{GS}_0
\Bigr),
\label{eq:H_0_s_discrete}
\\
&
\widetilde{H}_0^P  
 =  \! \sum_{\sigma}  \sum_{m}   \varepsilon_m \,
\Bigl( 
\widetilde{p}^{\dagger}_{\varepsilon_m,\sigma} 
 \widetilde{p}^{\phantom{\dagger}}_{\varepsilon_m,\sigma}
-
\langle
\widetilde{p}^{\dagger}_{\varepsilon_m,\sigma} 
 \widetilde{p}^{\phantom{\dagger}}_{\varepsilon_m,\sigma}
\rangle^{\mathrm{GS}}_0 \Bigr), 
\label{eq:H_0_p_discrete}
\end{align}
with the normalization 
 $\{ \widetilde{\alpha}^{\phantom{\dagger}}_{\varepsilon_m,\sigma}, 
 \widetilde{\alpha}^{\dagger}_{\varepsilon_{m'},\sigma'}
 \} = \delta_{mm'}\delta_{\sigma\sigma'}$, and 
 $\{ \widetilde{p}^{\phantom{\dagger}}_{\varepsilon_m,\sigma}, 
 \widetilde{p}^{\dagger}_{\varepsilon_{m'},\sigma'}
 \} = \delta_{mm'}\delta_{\sigma\sigma'}$. 
The constant term  $\langle \cdots \rangle^\mathrm{GS}_0$, 
which denotes the noninteracting ground-state average in equilibrium, 
is subtracted so as to make 
the lowest energy of $\widetilde{H}_0^S$ 
and that of $\widetilde{H}_0^P$ zero, 
\begin{align}
\!\!
\langle
\widetilde{\alpha}^{\dagger}_{\varepsilon_m,\sigma} 
 \widetilde{\alpha}^{\phantom{\dagger}}_{\varepsilon_m,\sigma}
\rangle^\mathrm{GS}_0
=\langle
\widetilde{p}^{\dagger}_{\varepsilon_m,\sigma} 
 \widetilde{p}^{\phantom{\dagger}}_{\varepsilon_m,\sigma}
\rangle^\mathrm{GS}_0
= \left\{
\!\!
\begin{array}{cl}
1, &   \varepsilon_m < 0 ,
\\ 
0,      &     \varepsilon_m > 0 .
\end{array}
\right. 
\end{align}
Specifically, 
we assume a linear discretization 
with a finite level spacing $\delta \varepsilon$, 
such that $\varepsilon_m = (m+1/2) \,\delta \varepsilon$ 
for $m=0,\pm 1, \pm 2, \ldots$. 
The continuous spectrum can be recovered 
in the limit of  $\delta \varepsilon \to 0$ with 
 $\alpha_{\varepsilon,\sigma}^{\phantom{\dagger}} =
\widetilde{\alpha}_{\varepsilon_m,\sigma}^{\phantom{\dagger}} / 
\sqrt{\delta \varepsilon}$ 
and 
$ p_{\varepsilon,\sigma}^{\phantom{\dagger}} =
\widetilde{p}_{\varepsilon_m,\sigma}^{\phantom{\dagger}}/ 
 \sqrt{\delta \varepsilon}$.
The discretized operator 
is also introduced 
for the propagating states 
$\widetilde{\gamma}^{\phantom{\dagger}}_{\varepsilon_m,\nu\sigma}$
using  Eq.\ (\ref{eq:scattering_state}), 
and then $\widetilde{H}_0$ can also be written in the form
\begin{align}
\widetilde{H}_0  
 = \!
\sum_{\nu} 
\sum_{\sigma} 
 \sum_{m}   \varepsilon_m \,
\Bigl( 
\widetilde{\gamma}^{\dagger}_{\varepsilon_m,\nu\sigma} 
 \widetilde{\gamma}^{\phantom{\dagger}}_{\varepsilon_m,\nu\sigma}
-
\langle
\widetilde{\gamma}^{\dagger}_{\varepsilon_m,\nu\sigma} 
 \widetilde{\gamma}^{\phantom{\dagger}}_{\varepsilon_m,\nu\sigma}
\rangle^\mathrm{GS}_0
\Bigr).
\end{align}
In this discretization,
each discrete level $\varepsilon_m$ preserves 
the two-fold degeneracy due to 
the left-going and right-going scattering states.
This degeneracy is essential to describe 
the current-carrying state with the density matrix.

The trace over the Hilbert space 
can be carried out explicitly for the discretized model. 
The normalization factor 
$\widetilde{\Xi}_0 =
\mbox{Tr}\, 
 e^{ -\beta (\widetilde{H}_0- \widetilde{Y}_0 )}$,
where $\widetilde{Y}_0$ is the discretized version 
of Eq.\ (\ref{eq:Y_0}),
% (\ref{eq:H0_Jost2}),
can be calculated as
\begin{align}
\widetilde{\Xi}_0=& \, 
\prod_{\nu=L,R}
\prod_{\sigma}
\prod_{\varepsilon_m} 
\left[1+e^{-\beta(\varepsilon_m-\mu_{\nu})\, 
\mathrm{sgn}\,\varepsilon_m} \right] \;.
% \prod_{\varepsilon'<0} 
% \left[1+e^{\beta(\varepsilon'-\mu_{\nu})} \right]  .
\label{eq:xi_0_expression}
\end{align}
Here, we have used a property $e^{\beta(\mu_L+\mu_R)}=1$ 
following from $\mu_L=-\mu_R=eV/2$.
The corresponding {\em free energy} that is defined by 
$\widetilde{\Xi}_0=e^{-\beta F_0}$ takes the form
\begin{align}
\beta  F_0 = - 
\frac{1}{\delta \varepsilon}
\sum_{\nu=L,R}
\!\sum_{\sigma} 
\!\int \!\! d\varepsilon \, \ln 
\left[1+e^{-\beta(\varepsilon-\mu_{\nu})\, 
\mathrm{sgn}\,\varepsilon } 
\right]. 
\end{align}
In the continuum limit $\delta \varepsilon\to 0$,
the factor $1/\delta \varepsilon$ should be replaced 
by the Dirac $\delta$-function $\delta(0)$. 
It appears because the energy of the whole system is 
proportional to the system size $N$. 
For instance, the level spacing is given by 
$\delta \varepsilon = 2D/N$ for a flat band with a half width $D$.
Therefore the level spacing $\delta \varepsilon$ 
has the physical meanings for the averages of order $N$, 
although we have started from the continuous model 
in order to capture properly 
the two-fold degeneracy in the scattering states.

The density matrix can also be calculated as,  
\begin{align}
\widetilde{\rho}_0
\,\equiv& \ 
\frac{e^{-\beta(\widetilde{H}_0 -\widetilde{Y}_0)}}
{\mbox{Tr}\,e^{-\beta(\widetilde{H}_0 -\widetilde{Y}_0)}}
\\
=& 
\prod_{\nu=L,R}\!
\prod_{\sigma}\prod_{\varepsilon} 
%\left[
\frac{
1+\left[\,e^{-\beta(\varepsilon-\mu_{\nu})}-1\,\right]
\widetilde{\gamma}_{\varepsilon,\nu\sigma}^{\dagger} 
\widetilde{\gamma}_{\varepsilon,\nu\sigma}^{\phantom{\dagger}}
}{1+e^{-\beta (\varepsilon-\mu_{\nu})}}
%\right] 
\;.
\label{eq:rho_0_expression}
\end{align}
Here,  
we have suppressed the suffix $m$ of 
the discrete frequency $\varepsilon_m$, for simplicity.
Equation (\ref{eq:rho_0_expression}) 
can be rewritten in terms 
of $\widetilde{\alpha}_{\varepsilon,\sigma}^{\phantom{\dagger}}$
and $\widetilde{p}_{\varepsilon,\sigma}^{\phantom{\dagger}}$, by
substituting Eq.\  (\ref{eq:scattering_state}), as
\begin{align}
& 
\!\!\!\!\!\!
\widetilde{\rho}_0
% \frac{e^{-\beta(\widetilde{H}_0 -\widetilde{Y}_0)}}
% {\mbox{Tr}\,e^{-\beta(\widetilde{H}_0 -\widetilde{Y}_0)}} 
% e^{-\beta({\cal H}_0 -Y_0)} 
% \nonumber 
% \\
=
\prod_{\sigma}\prod_{\varepsilon} 
\bigl(1-f_L\bigr)
\bigl(1-f_R\bigr)
\biggl[ \  1\ + 
\nonumber 
\\
&
 + 
\left( 
\frac{v_L^2}{v_L^2+v_R^2}
\frac{f_L}{1-f_L} 
+
\frac{v_R^2}{v_L^2+v_R^2}
\frac{f_R}{1-f_R} -1
\right)
\widetilde{\alpha}_{\varepsilon\sigma}^{\dagger} 
\widetilde{\alpha}_{\varepsilon\sigma}^{\phantom{\dagger}}
\nonumber
\\
& 
 + 
\left( 
\frac{v_R^2}{v_L^2+v_R^2}
\frac{f_L}{1-f_L} 
+
\frac{v_L^2}{v_L^2+v_R^2}
\frac{f_R}{1-f_R} -1
\right)
\widetilde{p}_{\varepsilon\sigma}^{\dagger} 
\widetilde{p}_{\varepsilon\sigma}^{\phantom{\dagger}}
\nonumber\\
&
-\frac{v_Lv_R}{v_L^2+v_R^2}\frac{f_L-f_R}{(1-f_L)(1-f_R)}
\left(
\widetilde{\alpha}_{\varepsilon\sigma}^{\dagger} 
\widetilde{p}_{\varepsilon\sigma}^{\phantom{\dagger}}
+
\widetilde{p}_{\varepsilon\sigma}^{\dagger} 
\widetilde{\alpha}_{\varepsilon\sigma}^{\phantom{\dagger}}
\right)
\nonumber\\
&
+ 
\frac{(1-2f_L)(1-2f_R)}{(1-f_L)(1-f_R)}\,
\widetilde{\alpha}_{\varepsilon\sigma}^{\dagger} 
\widetilde{\alpha}_{\varepsilon\sigma}^{\phantom{\dagger}}
\,
\widetilde{p}_{\varepsilon\sigma}^{\dagger} 
\widetilde{p}_{\varepsilon\sigma}^{\phantom{\dagger}}
\,\biggr] .
\label{eq:rho_0_sp}
\end{align}
Furthermore, the partial trace over 
the $p$-wave part can be carried out 
to yield the RDM for the $s$-wave subspace 
\begin{align}
 \mbox{Tr}_{P}\,  
 \widetilde{\rho}_0  
 =& 
  \prod_{\sigma}\prod_{\varepsilon} 
\left(1-f_\mathrm{eff}\right)
  \left[ 1 +
\frac{2f_\mathrm{eff}-1}{1-f_\mathrm{eff}}
\, \widetilde{\alpha}_{\varepsilon\sigma}^{\dagger} 
  \widetilde{\alpha}_{\varepsilon\sigma}^{\phantom{\dagger}}
  \right]. 
\label{eq:rho_s_0_primitive}
\end{align}
It can also be expressed in an exponential form,  
\begin{align}
 &
\widehat{\rho}_0^S  \equiv    
 \mbox{Tr}_{P}\,  
 \widetilde{\rho}_0  
\ = 
\frac{e^{-K_0^S}}{\mbox{Tr}_S\ e^{-K_0^S}}\;,
\label{eq:rho_s_0_T}
\\
&
\!\!\!
K_0^S \equiv \sum_{\sigma} \sum_{\varepsilon}
 \frac{\varepsilon}{\mathcal{T}_{\varepsilon}} \,
\widetilde{\alpha}_{\varepsilon\sigma}^{\dagger} 
\widetilde{\alpha}_{\varepsilon\sigma}^{\phantom{\dagger}}
\xrightarrow[\ \delta \varepsilon \to 0\ ]{} 
\sum_{\sigma} 
\int\! d\varepsilon\, 
 \frac{\varepsilon}{\mathcal{T}_{\varepsilon}} \,
\alpha_{\varepsilon\sigma}^{\dagger} 
\alpha_{\varepsilon\sigma}^{\phantom{\dagger}} ,
\label{eq:K_s}
\\
&\frac{1}{\mathcal{T}_{\varepsilon}}
\equiv
 \frac{1}{\varepsilon} 
\ln \! \left[
\frac{1-f_\mathrm{eff}(\varepsilon)}
{f_\mathrm{eff}(\varepsilon)}
 \right] . 
\label{eq:T_eff}
\end{align}

\begin{figure}[t]
 \includegraphics[width= 0.78\linewidth]{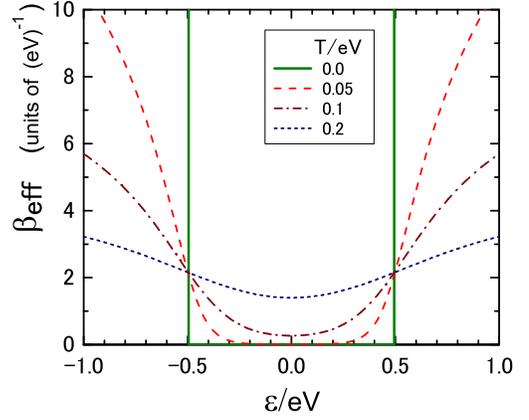}
\caption{
(color online). 
Parameter  $\beta_\mathrm{eff} 
\equiv 1/\mathcal{T}_{\varepsilon}$, 
which determines the RDM as Eq.\ (\ref{eq:K_s}), 
 is plotted as a function of $\varepsilon$ for 
several temperatures $T/(eV)=0.0,\, 0.05,\, 0.1$ and $0.2$, 
in the symmetric coupling case $\Gamma_L=\Gamma_R$.
}
\label{fig:beff} 
\end{figure}

Then, the average of an operator 
$\widehat{\cal O}_S = 
\widehat{\cal O}_S(\alpha,\alpha^{\dagger})$, which  
is an arbitrary function of 
 $\alpha^{\phantom{\dagger}}_{\varepsilon,\sigma}$ 
and $\alpha^{\dagger}_{\varepsilon,\sigma}$,  
can be calculated in the $s$-wave subspace 
using the RDM,
\begin{align} 
&\langle 
   \widehat{\cal O}_S 
\rangle_0
\equiv
 \mbox{Tr}_S^{\phantom{|}} 
 \mbox{Tr}_P^{\phantom{|}} \!
\left[ 
 \widehat{\rho}_{0}^{\phantom{S}} 
\widehat{\cal O}_S 
\right]
=
 \mbox{Tr}_S^{\phantom{|}} \!
\left[ 
 \widehat{\rho}_{0}^{S}\,
\widehat{\cal O}_S 
\right]
.
\label{eq:with_rho_s_0}
\end{align}
Note that the impurity level 
has been included in the $s$-wave part.
It is straight forward to confirm 
that $\langle \alpha^{\dagger}_{\varepsilon,\sigma} 
 \alpha^{\phantom{\dagger}}_{\varepsilon',\sigma'}
\rangle_0$ given in Eq.\ (\ref{eq:f_eff}) can be 
reproduced from Eq.\ (\ref{eq:with_rho_s_0}),
and then the function $f_\mathrm{eff}(\varepsilon)$ 
defined in Eq.\ (\ref{eq:f_eff_def}) 
can  be expressed in terms of $\mathcal{T}_{\varepsilon}$ as
\begin{equation}
f_\mathrm{eff}(\varepsilon) = 
\frac{1}{e^{\varepsilon/\mathcal{T}_{\varepsilon}} +1} \;.
\end{equation}
The bias dependence appears in the $s$-wave subspace 
through this energy-dependent parameter 
$\mathcal{T}_{\varepsilon}$.
% which may be called {\em effective temperature\/}.
The inverse of this parameter,  
 $\beta_\mathrm{eff} 
\equiv 1/\mathcal{T}_{\varepsilon}$, is plotted 
in Fig.\ \ref{fig:beff} 
as a function of $\varepsilon$ for several 
values of real temperature $T$ 
in the symmetric coupling $\Gamma_L=\Gamma_R$ case.
The energy dependence of $\mathcal{T}_{\varepsilon}$
becomes weak with increasing $T$.
At zero temperature, it takes the form,
\begin{align}
 \mathcal{T}_{\varepsilon} \to 
 \left\{
\!
\begin{array}{cl}
\infty, &  \ \  |\varepsilon| < eV/2 , 
\\ 
0,      &   \ \   |\varepsilon| > eV/2 . 
\end{array}
\right. 
\end{align}
Namely, $\mathcal{T}_{\varepsilon}$ 
reaches infinity inside the energy window $|\varepsilon| < eV/2$, 
while it remain zero on the outside. 
For this reason,  
in the limit of $eV \to \infty$  
the nonequilibrium Green's functions at the impurity site 
 coincide with the equilibrium ones 
for the high-temperature limit $T\to \infty$.\cite{OguriB}

The partial trace over the $p$-wave part 
can be carried out also for $H_U \neq 0$,
as the Coulomb interaction affects 
only the $s$-wave part of the Hilbert space.
Therefore, the interacting RDM 
 $\widehat{\rho}_{\phantom{0}}^{S}$  evolves adiabatically 
from the noninteracting RDM $\widehat{\rho}_0^{S}$ as
\begin{align} 
\widehat{\rho}_{\phantom{0}}^{S}
=& \,
 S(0,-\infty) \,\widehat{\rho}_0^{S}
   S(-\infty,0)  \;,
   \label{eq:rho_s}
\\ 
\langle 
   \widehat{\cal O}_S 
 \rangle
 \equiv& \,
\mbox{Tr}_S^{\phantom{|}} 
\mbox{Tr}_P^{\phantom{|}}\! 
\left[\,
 \widehat{\rho}  \  \widehat{\cal O}_S  \right]  
=
 \mbox{Tr}_S^{\phantom{|}}\! \left[\,
 \widehat{\rho}_{\phantom{0}}^{S}
 \, \widehat{\cal O}_S  \right]  
 \;. 
 \label{eq:neq_average_u2}
\end{align}
It enables us to treat the bias voltage 
within the $s$-wave part of the Hilbert space.
Namely, one can start with the single-channel Anderson model $H^S$ 
in the continuous form, Eq.\ (\ref{eq:H_s}).
Then, the nonequilibrium distribution can be taken into account 
through the initial RDM $\widehat{\rho}_{0}^{S}$,
which evolves to the interacting one $\widehat{\rho}_{\phantom{0}}^{S}$. 
For the single-channel model, 
it is not necessary to use the same discretization 
scheme that was introduced 
to obtain Eq.\ (\ref{eq:rho_s_0_primitive}), 
because  $\widehat{\rho}_{0}^{S}$ 
has a well-defined continuum form as Eq.\ (\ref{eq:K_s}). 
In addition to these points, 
the correlation functions containing 
a $p$-wave operator can also be calculated 
with the single-channel model as shown in Appendix \ref{sec:RDM_for_p}.

\section{Mixed-state aspect for $\Gamma_L=\Gamma_R$}
\label{sec:mixed-state}

As we see in Fig.\ \ref{fig:beff},
effects of the bias voltage are pronounced 
significantly at zero temperature, 
 especially when the system has an inversion symmetry $\Gamma_L=\Gamma_R$. 
We consider in the following the properties of 
the RDM in this particular case.

At zero temperature $T=0$   
the noninteracting density matrix 
$\widehat{\rho}_0^{\phantom{0}}$ 
for the whole system can be constructed as 
a pure state $\left|  \Phi_{\rm neq} \right \rangle$, 
in which the scattering states incident  
from left and right are filled up to 
$\mu_L=eV/2$ and $\mu_R=-eV/2$, respectively,
\begin{align}
&\left|  \Phi_{\rm neq} \right \rangle
= 
\prod_{\varepsilon=0}^{\frac{eV}{2}} 
\prod_{\varepsilon'=-\frac{eV}{2}}^{0} 
\prod_{\sigma} 
\widetilde{\gamma}_{\varepsilon,L\sigma}^{\dagger}
\widetilde{\gamma}_{\varepsilon',R\sigma}^{\phantom{\dagger}}
\left|  \Phi_{0} \right \rangle \;,
\label{eq:Phi_0_neq}
\\
&\left| \Phi_{0} \right \rangle 
\equiv
\prod_{\varepsilon<0}\prod_{\sigma} 
\widetilde{\gamma}_{\varepsilon,L\sigma}^{\dagger}
\widetilde{\gamma}_{\varepsilon,R\sigma}^{\dagger}
\left|0\right\rangle =
 \prod_{\varepsilon<0}\prod_{\sigma} 
\widetilde{\alpha}_{\varepsilon\sigma}^{\dagger}
\widetilde{p}_{\varepsilon\sigma}^{\dagger}
 \left|0\right\rangle .
\label{eq:Phi_0}
\end{align}
Here, $\left|0\right\rangle$ is the vacuum state.
The noninteracting ground state 
 $\left|  \Phi_{0}^{} \right \rangle$ can be 
decomposed into the Fermi sea of the $s$-wave part 
$\left| \Phi_{0}^{S}\right \rangle$ and that of 
the $p$-wave part $\left| \Phi_{0}^{P}\right \rangle$ as
$\left|  \Phi_{0}^{} \right \rangle =
\left| \Phi_{0}^{S} \right \rangle 
\left| \Phi_{0}^{P} \right \rangle$. 
Therefore, Eq.\ (\ref{eq:Phi_0_neq}) can be 
rewritten, using Eq.\ (\ref{eq:scattering_state}) 
for $v_L=v_R$, as
\begin{align}
\left|  \Phi_{\rm neq} \right \rangle
&=\, 
\prod_{\varepsilon=0}^{\frac{eV}{2}} 
\prod_{\varepsilon'=-\frac{eV}{2}}^{0} 
\prod_{\sigma} 
\frac{(\widetilde{\alpha}_{\varepsilon\sigma}^{\dagger}
- \widetilde{p}_{\varepsilon\sigma}^{\dagger})}{\sqrt{2}}
\frac{(\widetilde{\alpha}_{\varepsilon'\sigma}^{\phantom{\dagger}}
 + \widetilde{p}_{\varepsilon'\sigma}^{\phantom{\dagger}})}{\sqrt{2}}
\,\left|  \Phi_{0} \right \rangle
\nonumber \\
&=\, 
\frac{1}{\sqrt{Z_0}}\,{\sum_{i \in \mathcal{W} }}'
\left| \Phi_{i}^{S} \right \rangle 
\left| \Phi_{i}^{P} \right \rangle 
\;.
\label{rho_0_mixed}
\end{align}
Here, $\left| \Phi_{i}^{S} \right \rangle$ 
and $\left| \Phi_{i}^{P} \right \rangle$ are
the excited states of the $s$-wave part and that of the $p$-wave part, 
respectively: the sign caused by exchanges of  
 the fermion operators is absorbed into 
 the global phase of each many-particle eigenstate 
in the definition.
The primed sum for $i$ runs 
over a set $\mathcal{W}$, which consists of 
all the possible excited states that can be 
generated from $\left|  \Phi_{0}^{} \right \rangle$
by changing the occupation of the one-particle states 
inside the biased energy region $|\varepsilon|<eV/2$. 
The factor $Z_0$ corresponds to the dimension of $\mathcal{W}$.

The pure-state average of an arbitrary operator 
$\widehat{\cal O}_S = 
\widehat{\cal O}_S(\alpha,\alpha^{\dagger})$ 
that is defined in the $s$-wave space can be calculated, 
 carrying out the summation over the $p$-wave part, as
 \begin{align}
& \left\langle  \Phi_{\rm neq} \right |
 \widehat{\cal O}_S 
\left|  \Phi_{\rm neq} \right \rangle
=
\frac{1}{Z_0}\,{\sum_{ij\in\mathcal{W}}}'
\!
\left \langle \Phi_{i}^{P} \right| \!\!  
\left\langle \Phi_{i}^{S} \right | 
 \widehat{\cal O}_S 
\left| \Phi_{j}^{S} \right \rangle \!\!
\left| \Phi_{j}^{P} \right \rangle 
\nonumber
\\
& \qquad \quad \ \  
= 
\frac{1}{Z_0}\,{\sum_{i\in\mathcal{W}}}'
\!\left\langle \Phi_{i}^{S} \right | 
 \widehat{\cal O}_S 
\left| \Phi_{i}^{S} \right \rangle 
= 
\mbox{Tr}_S^{\phantom{|}}\! \left[\,
 \widehat{\rho}_0^{S}  
\, \widehat{\cal O}_S 
 \right] .
\label{eq:mixed_s_first}
\end{align}
Therefore in this case, 
the noninteracting RDM takes a mixed-state 
form  $\widehat{\rho}_{0}^{S} 
=
\sum_n \! 
\left|\Phi_{n}^{S}\right \rangle  w_{0,n}^{S} \!  
\left\langle \Phi_{n}^{S}\right|$ with  
\begin{align}
% \ \ \, 
w_{0,n}^{S} =  
\left\{\!
\begin{array}{cl}
1/Z_0, &\quad  n \in \mathcal{W} ,   \\ 
0,     &\quad n \notin \mathcal{W} .
\end{array}
\right. 
\label{eq:w_n^S}
\end{align}
The summation with respect to $n$, 
or $\mbox{Tr}_S^{\phantom{|}}[\cdots]$, 
run over the whole region of the $s$-wave subspace.
The statistical weight $w_{0,n}^{S}$ takes just two values 
$0$ or $1/Z_0$. However, it 
is not a simple function of the many-particle 
energy eigenvalue $E_{0,n}^S$ 
of $\widetilde{H}_0^S\left|\Phi_{n}^{S}\right \rangle 
= E_{0,n}^S\left|\Phi_{n}^{S}\right\rangle$.

There is another important relation between   
$\left| \Phi_{i}^{S} \right \rangle$ and 
$\left| \Phi_{i}^{P} \right \rangle$, 
as these states are generated 
simultaneously in Eq.\ (\ref{rho_0_mixed}).
If a one-particle level $\varepsilon$ 
in the biased energy region $|\varepsilon| < eV/2$ 
is being occupied by an $s$-wave electron  
$\widetilde{\alpha}_{\varepsilon\sigma}^{\dagger}$  in  
 $\left| \Phi_{i}^{S} \right \rangle$, 
then the corresponding level has to be unoccupied 
in the $p$-wave part $\left| \Phi_{i}^{P} \right \rangle$, 
or vice versa. Therefore, there exists a pair of states,  
$i$ and $i'$, that show a relation  
 $\widetilde{\alpha}_{\varepsilon\sigma}^{\phantom{\dagger}}
\left| \Phi_{i}^{S} \right \rangle 
\!\left| \Phi_{i}^{P} \right \rangle 
= -
\,\widetilde{p}_{\varepsilon\sigma}^{\phantom{\dagger}}
\left| \Phi_{i'}^{S} \right \rangle 
\!\left| \Phi_{i'}^{P} \right \rangle 
$ for  $|\varepsilon| < eV/2$. 
Here, the minus sign is caused by the order 
of the operators for the one-particle level $\varepsilon$.
Using this property, 
a $p$-wave operator appearing 
in an average can be replaced by an $s$-wave operator as
\begin{align}
&
\!\!\!\!\!\!
\left\langle  \Phi_{\rm neq} \right |
\widehat{\cal O}_S \, p_{\varepsilon\sigma}
\! \left|  \Phi_{\rm neq} \right \rangle
= 
\frac{1}{Z_0} 
{\sum_{ij\in\mathcal{W}}}'
\!
\left\langle \Phi_{i}^{P} \right |  
\!\!
\left\langle \Phi_{i}^{S} \right |  
\widehat{\cal O}_S \, 
p_{\varepsilon\sigma} \! 
\left| \Phi_{j}^{S} \right \rangle 
\!\!
\left| \Phi_{j}^{P} \right \rangle 
\nonumber
\\
& \qquad \qquad \quad   
=  
-\,
\left[ 
f_L(\varepsilon) - f_R(\varepsilon) 
\right]\, 
\mbox{Tr}_S^{\phantom{|}}\! \left[
\widehat{\rho}_0^{S}  
\, \widehat{\cal O}_S \,
\alpha_{\varepsilon\sigma}
 \right] . 
\label{eq:with_p}
\end{align}
 Similar relation holds also 
for the asymmetric coupling $\Gamma_L \neq \Gamma_R$ 
and at finite temperatures as shown in Appendix \ref{sec:RDM_for_p}.

The interacting  density matrix at $T=0$ 
is also constructed as a pure state 
with the wavefunction
\begin{align}
\left|  \Psi_{\rm neq} \right \rangle
\equiv 
 S(0,-\infty) \left|  \Phi_{\rm neq} \right \rangle
=
\frac{1}{\sqrt{Z_0}}
{\sum_{i\in \mathcal{W}}}' 
\left|\Psi_{i}^{S}\right\rangle 
\left|\Phi_{i}^{P}\right\rangle \;.
\label{eq:rho_mixed}
\end{align}
Here, $\left|\Psi_{n}^{S}\right\rangle 
\equiv S(0,-\infty) \left|\Phi_{n}^{S}\right\rangle$, 
and it is an interacting eigenstate,\cite{GML}  
 $(\widetilde{H}_0^S+H_U)\left|  \Psi_{n}^{S} \right \rangle 
= E_n^{S} \left|  \Psi_{n}^{S} \right \rangle$.
The wave function for the $p$-wave part keeps 
the noninteracting form, because $S(0,-\infty)$ 
commutes with the $p$-wave operators. 
Therefore, 
the summation over the $p$-wave part can be carried out 
in the same way as that in Eq.\ (\ref{eq:mixed_s_first}),
and the interacting average for the operator
$\widehat{\cal O}_S$ can be written in the form,   
\begin{align}
\left\langle  \Psi_{\rm neq} \right |
 \widehat{\cal O}_S 
\left|  \Psi_{\rm neq} \right \rangle
& 
 = 
\frac{1}{Z_0}
{\sum_{i\in \mathcal{W}}}'  \left\langle \Psi_{i}^{S}\right| 
\widehat{\cal O}_S \left|\Psi_{i}^{S} \right\rangle 
\nonumber \\
&=
\sum_n  w_{0,n}^{S} \!\left\langle \Psi_{n}^{S}\right| 
\widehat{\cal O}_S \left|\Psi_{n}^{S} \right\rangle . 
\label{eq:neq_average_u}
\end{align}
The interacting RDM in this case also takes a simple form  
$
\widehat{\rho}_{\phantom{0}}^{S}
=
\sum_n
\left|\Psi_{n}^{S}\right\rangle  w_{0,n}^{S} 
\!\left\langle \Psi_{n}^{S}\right| 
$.
 Equation (\ref{eq:neq_average_u})  
gives us much insight into the steady state. 
It will provide, obviously, 
an interacting equilibrium average if we replace $w_{0,n}^{S}$
by $\rho_{\mathrm{eq},n}^S \equiv 
e^{-\beta E_n^S}\!/\! \sum_n \! e^{-\beta E_n^S}$. 
Therefore, some of the nonequilibrium properties 
could be inferred from the difference 
between the statistical weights $w_{0,n}^{S}$ and $\rho_{\mathrm{eq},n}^S$.

As Eq.\ (\ref{eq:neq_average_u}) is being expressed 
in terms of the interacting eigenstates,  
Hamiltonian-based nonperturbative approaches developed for equilibrium, 
such as exact diagonalization and NRG, can be applied to 
the matrix element $\left\langle \Psi_{n}^{S}\right| 
\!\widehat{\cal O}_S \!\left|\Psi_{n}^{S} \right\rangle$. 
Furthermore, the bias voltage appears just through $w_{0,n}^{S}$,  
which keeps the noninteracting form given in Eq.\ (\ref{eq:w_n^S}). 
Thus, what is necessary 
to carry out the summation  with 
respect to  $n$ in Eq.\ (\ref{eq:neq_average_u}) is 
the information about
whether or not each interacting 
eigenstate $\left| \Psi_{n}^{S} \right \rangle$ 
belongs to the set $\mathcal{W}$ in the limit of $H_U \to 0$. 
Namely, the {\em correspondence\/} between 
  $\left|\Psi_{i}^{S}\right\rangle$ and its  
 noninteracting counterpart $\left|\Phi_{i}^{S}\right\rangle$ 
is an additional issue that one has to solve   
in the nonequilibrium case. 
At low energies, 
the {\em correspondence\/} can be treated correctly 
with the quasi-particles of 
a local Fermi liquid.\cite{Hewson,KNG,OguriA,FujiiUeda,GogolinKomnik,HB}

The evolution of the wavefunctions 
 is described formally with the time-dependent 
operator $S(0,-\infty)$.
Alternatively, 
it can be expressed in a time-independent form 
as that in a stationary-state scattering theory,\cite{GML,GMGB}
and several versions of the time-independent approaches 
have been examined 
with some different points of view.
\cite{SchillerHershfield,KSL,KNG,Han,MehtaAndrei} 
In our description,
the adiabatic evolution of the interacting eigenstates 
from the noninteracting ones could also be traced directly 
in the Hilbert space of a discretized version of 
the single-channel model, 
by changing the coupling constant $U$ continuously.

\section{Energy distribution in the  mixed states }
\label{sec:average}

We consider in the following the mixed-sate properties 
for the symmetric coupling $\Gamma_L=\Gamma_R$ further at $T=0$.
The pure state $\left| \Phi_{\rm neq} \right \rangle$
given in Eq.\ (\ref{eq:Phi_0_neq})
is an eigenstate, 
$\widetilde{H}_0 \left| \Phi_{\rm neq} \right \rangle =
E_0^\mathrm{neq}
\left| \Phi_{\rm neq} \right \rangle$, of the noninteracting Hamiltonian
with an excitation energy
\begin{align}
E_0^\mathrm{neq} 
\, = \, 2 \! 
\sum_{\varepsilon_m = -eV/2}^{eV/2} |\varepsilon_m|
\xrightarrow[\ \delta \varepsilon \to 0 \ ]{} 
\frac{\left(eV\right)^2}{2\,\delta \varepsilon}
\;.
\label{eq:pure_state_energy}
\end{align}
Here, we retained only the dominant term 
that is proportional to the system size of order $N$. 
As mentioned in Sec.\ \ref{sec:RDM},  
the level spacing is given by  
 $\delta \varepsilon = 2D/N$ for a flat band with a half width $D$. 
The feature of the pure state reflects 
in the $s$-wave subspace through the RDM.
The mean excitation energy 
in the subspace is equal to  
a half of the one in Eq.\ (\ref{eq:pure_state_energy}),  
$\overline{E}_0^S \equiv  
\mbox{Tr}_S^{\phantom{|}} \bigl[
\widehat{\rho}_0^S \widetilde{H}_0^S
\bigr] = E_0^\mathrm{neq}/2$. 
Note that the contribution of the impurity level is 
of order $1$, namely a $1/N$ portion of the total energy.

\begin{figure}[t]
 \includegraphics[width= 0.78\linewidth]{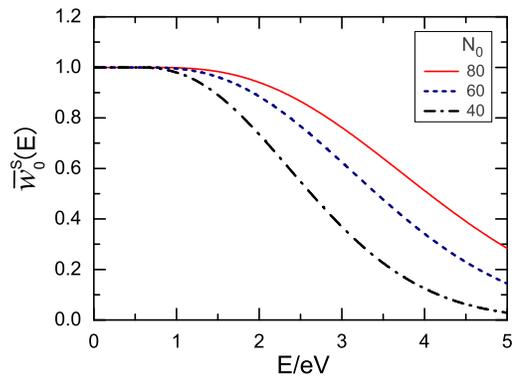}
\caption{
(color online).
Averaged statistical weight
 $\overline{w}_0^{S}(E) \equiv W(E)\,/\,\Omega(E)$ 
 vs  many-particle excitation energy $E$, 
where $N_0 = eV/\delta\varepsilon$.
}
\label{fig:wapp} 
\end{figure}

In order to clarify how the nonequilibrium weight $w_{0,n}^{S}$ 
distributes in the many-particle Hilbert space consisting 
of the eigenstates $\left| \Phi_{n}^{S} \right \rangle$, 
we examine an averaged weight $\overline{w}_0^{S}(E)$ 
over an equienergy surface; 
\begin{align}
 & \overline{w}_0^{S}(E) \, \equiv \,
 W(E)\,/\,\Omega(E) \;, 
 \\
&
\!\!\!
W(E) = \!
 {\sum_{i\in\mathcal{W}}}' \! 
  \delta (E-E_{0,i}^{S}) , \ \ 
 \Omega(E) =\!\sum_n \delta (E-E_{0,n}^{S}). 
 \label{eq:MB_DOS}
\end{align}
It is normalized as $Z_0 = \int \! dE\  \overline{w}_0^{S}(E) \Omega(E)$,
and may be used as 
$
\left\langle  \Psi_{\rm neq} \right |
 \! \widehat{\cal O}_S \!
\left|  \Psi_{\rm neq} \right \rangle 
 \approx \!
\sum_n  \overline{w}_0^{S}(E_n^S) 
\left\langle \Psi_{n}^{S}\right|
\! \widehat{\cal O}_S \!
\left|\Psi_{n}^{S} \right\rangle \!/Z_0
$ for a rough approximation.    
We calculate $\overline{w}_0^{S}(E)$ for 
several finite $\delta \varepsilon$,
and the results are shown in Fig.\ \ref{fig:wapp}:
the parameter $N_0 \equiv eV/\delta \varepsilon$ 
corresponds to the number of one-particle states 
in the energy window $|\varepsilon|<eV/2$, 
and thus $Z_0 = 4^{N_0}$. 
The averaged weight $\overline{w}_0^{S}(E)$ becomes a constant for small $E$,
as all the many-particle excited states for 
$0<E<eV/2$ belong to the set $\mathcal{W}$  defined 
in Eq.\ (\ref{rho_0_mixed}). 
This flat $E$-dependence is quite different from  
the exponential dependence of the Boltzmann factor $e^{-\beta E}$ in 
the thermal equilibrium. 
 We also see that $\overline{w}_0^{S}(E)$ penetrates into 
the higher energy region over the bias energy $eV$.
It spreads out, approximately, 
up to $E \sim c\, \overline{E}_0^S$ with 
a numerical constant $c$ of order $10^{-1}$.
Note that the mean excitation energy is proportional to $N_0$ 
as $\overline{E}_0^S=N_0 eV/4$. 
We have also obtained the analytic expressions of 
 $\Omega(E)$ and $W(E)$ for small $\delta \varepsilon$ 
as shown in Appendix \ref{sec:averaged_w_details},  
and have checked that 
the same behavior could be seen for larger $N_0$
 assuming  a Gaussian form 
for $W(E)$ given in Eq.\ (\ref{eq:W_integral_app}), 
instead of the exact integration form Eq.\ (\ref{eq:W_integral}).

The feature of $\overline{w}_0^{S}(E)$ 
 seen in the $s$-wave subspace  
is caused by the fact that the bias voltage has been 
introduced not into the many-particle energy 
but into the one-particle states  
as the width of the energy window 
between the two chemical potentials $\mu_L$ and $\mu_R$. 
It is quite different from the way 
the temperature determines the equilibrium distribution $e^{- E/T}$,
in which $T$ scales directly the many-particle energy $E$. 
The original weight $w_{0,n}^{S}$ must have the similar features 
in the $n$ dependence. However,
whether or not the high-energy many-particle states 
give significant contributions to the nonequilibrium averages 
depends also on the properties of 
the operator $\widehat{\cal O}_S$.
For instance, the 
 noninteracting correlation functions, 
such as  $\langle \alpha^{\dagger}_{\varepsilon,\sigma} 
 \alpha^{\phantom{\dagger}}_{\varepsilon',\sigma'}
\rangle_0$ and $G^{<}_{0}(\varepsilon)$,  
can be reproduced from Eq.\ (\ref{eq:mixed_s_first}). 
 The feature of $w_{0,n}^{S}$ 
reflects in these correlation functions  
just through $f_\mathrm{eff}(\varepsilon)$. 
It should be regarded, however, 
those are the results that are determined by 
all the many-particle states belonging to the set $\mathcal{W}$.
Therefore, 
it is  necessary for calculating the interacting Green's  
functions with Eq.\ (\ref{eq:neq_average_u}) 
to take all these high-energy states into account. 
The Kondo resonance at finite 
bias voltages is being affected by these features of 
the nonequilibrium distribution.

\section{Summary}
\label{sec:summary}

  In conclusion, we have reexamined  
the nonequilibrium steady state for a quantum dot, 
checking out the statistical distribution   
in the many-particle Hilbert space. 
We have, first of all, discretized the conduction bands 
keeping the two-fold degeneracy 
due to the left-going and right-going scattering states.
This degeneracy is essential to describe the current-carrying 
state driven by the bias voltage.  
Then, by integrating out the $p$-wave part of the channel 
degrees of freedom, the noninteracting RDM  $\widehat{\rho}_0^S$
has been obtained explicitly as shown in Eq.\ (\ref{eq:rho_s_0_T}). 
It brings the information about the bias voltages 
to the remaining $s$-wave subspace,  
and determines the nonequilibrium  distribution. 
The interacting RDM $\widehat{\rho}^S$, 
which is defined in Eq.\ (\ref{eq:rho_s}),
 evolves adiabatically from  $\widehat{\rho}_0^S$.
Using the RDM, 
the steady-state averages Eq.\ (\ref{eq:neq_average_u2}) 
can be calculated, in principle, with the single-channel Anderson model.

We have studied further the properties of the RDM 
 at zero temperature for the symmetric coupling $\Gamma_L=\Gamma_R$.
In this case the nonequilibrium averages can be 
written in a Lehmann-representation form Eq.\ (\ref{eq:neq_average_u}) 
with a simple mixed-state distribution $w_{0,n}^{S}$
 given in Eq.\ (\ref{eq:w_n^S}).
It has also been deduced that the mixed states 
contain large portions of high-energy many-particle states.
Our formulation could be used for constructing 
Hamiltonian-based nonperturbative approaches in future.

\begin{acknowledgements}

I wish to thank Y. Nisikawa, J. Bauer, and A. C. Hewson  
for valuable discussions.  This work was supported 
by the Grant-in-Aid for Scientific Research from JSPS.

\end{acknowledgements}

\appendix

\section{RDM approach to the $p$-wave operator}
\label{sec:RDM_for_p}

The correlation functions 
which include a $p$-wave operator 
can be expressed, using Eq.\ (\ref{eq:rho_0_sp}), as
\begin{align} 
&
\langle 
   \widehat{\cal O}_S \,
 p^{\phantom{\dagger}}_{\varepsilon\sigma} 
\rangle_0
 = 
\lim_{\delta \varepsilon \to 0}
 \mbox{Tr}_S^{\phantom{|}}
 \mbox{Tr}_P^{\phantom{|}}
\! \left[\,
 \widetilde{\rho}_{0}
 \, \widehat{\cal O}_S  \,
\widetilde{p}^{\phantom{\dagger}}_{\varepsilon,\sigma} 
 \right]/\,\sqrt{\delta \varepsilon}  
\nonumber
\\
& \qquad 
=
- 
\frac{v_L^{\phantom{|}}v_R^{\phantom{|}}}
{v_R^2+v_L^2}
\frac{f_L(\varepsilon) -f_R(\varepsilon)}
{f_\mathrm{eff}(\varepsilon)}\, 
 \mbox{Tr}_S^{\phantom{|}} \!
\left[ 
 \widehat{\rho}_{0}^{S}\,
\widehat{\cal O}_S \,
\alpha^{\phantom{\dagger}}_{\varepsilon\sigma} 
\right]
\label{eq:with_p_T}
\end{align}
in the noninteracting case. 
For instance,
 Eq.\ (\ref{eq:alpha_p}) can be reproduced from Eq.\ (\ref{eq:with_p_T}) 
for $\widehat{\cal O}_S = \alpha_{\varepsilon'\sigma'}^{\dagger}$.
The interacting average can also be expressed in a similar form
\begin{align}
&\langle 
 \widehat{\cal O}_S \, p^{\phantom{\dagger}}_{\varepsilon\sigma} 
\rangle
=
- \frac{v_L^{\phantom{|}}v_R^{\phantom{|}}}
{v_R^2+v_L^2}
\frac{f_L(\varepsilon) -f_R(\varepsilon)}
{f_\mathrm{eff}(\varepsilon)}\, 
\nonumber \\
& \qquad \qquad \times
\mbox{Tr}_S \left[\widehat{\rho}_0^{S}\,  
S(-\infty,0)\,\widehat{\cal O}_S\, S(0,-\infty)
\,\alpha_{\varepsilon\sigma}^{\phantom{\dagger}}
 \right].
\end{align}

\section{$\Omega(E)$ and $W(E)$ for small $\delta \varepsilon$}
\label{sec:averaged_w_details}

We provide the integration forms of
$\Omega(E)$ and $W(E)$ for small
level spacing $\delta\varepsilon$.
To this end, we consider a grand partition function $\mathcal{G}(\tau,E_c)$
for noninteracting electrons in a flat band 
with the cut-off $E_c$. 
The grand partition function and many-particle 
density of states $F(E,E_c)$ are relating each other 
with the Laplace transform,  
\begin{align}
&
\mathcal{G}(\tau,E_c) 
=
\int_0^{\infty} \!\!dE\ e^{-\tau E}\,F(E,E_c) 
\;.
\label{eq:Laplace}
\end{align}
For small $\delta\varepsilon$,   
the left-hand side, $\mathcal{G}(\tau,E_c)$, is given by 
\begin{align}
&
 \prod_{\sigma}
\prod_{\lambda=e,h}
\prod_{\varepsilon=0}^{E_c}
\left(1+e^{-\varepsilon\tau}\right)
 \Rightarrow
\prod_{\sigma}
\frac{1+e^{-\tau E_c}}{2} 
\  e^{
\frac{\scriptstyle 2}{\scriptstyle \delta \varepsilon^{}} \,
X(\tau,E_c)
} ,
\\
&
\!\!\!
X(\tau,E_c)
\equiv 
\!\! \int_0^{E_c} \!\!\! d\varepsilon 
\, \ln\left(1+e^{-\varepsilon\tau}\right) 
= \frac{\phi(0)-\phi(\tau E_c)}{\tau} .
\end{align}
Here, $\phi(z) =  
\sum_{m=1}^{\infty} \frac{(-1)^{m-1}}{m^2} e^{-m z}$ and 
 $\phi(0)=\pi^2/12$.  
The label $\lambda$ represents 
 the contributions of the electron ($e$) and  hole ($h$) excitations. 
The function 
$F(E,E_c)$ is given as the inverse Laplace transform 
\begin{align}
F(E,E_c) = 
\!\! \int_{\tau_0-i\infty}^{\tau_0+i\infty}
\!\! \frac{d\tau}{2 \pi i}
\, e^{\tau E} \, 
\frac{\left(1+e^{-\tau E_c}\right)^2
}{4} 
e^{\frac{\scriptstyle 4}{\scriptstyle \delta \varepsilon} 
\frac{\scriptstyle \phi(0) -\phi(\tau E_c)}{\scriptstyle \tau}}.
\label{eq:inv_Laplace}
\end{align}

The function $\Omega(E) = F(E,\infty)$ is obtained 
taking the limit of $E_c \to \infty$:     
then $e^{-\tau E_c}\to 0$, $\phi(\tau E_c)\to 0$, 
and the integration can be carried out for $E>0$,   
\begin{align}
\Omega(E)\, &= \, \frac{1}{4}
\sqrt{\frac{4\phi(0)}{E\,\delta \varepsilon}}\ I_1
\!\left(2\sqrt{{4\phi(0)E}/{\delta \varepsilon}} \right) 
%%  + \delta(E)
 \:.
\label{eq:X_bessel}
\end{align}
Here, $I_1(x)$ is the modified 
Bessel function, which
 for large $x$   
takes the asymptotic form; $I_1(x) \simeq e^x/\sqrt{2\pi x}$.

The other function $W(E)=F(E,eV/2)$ is obtained for $E_c =eV/2$. 
Note that $W(E)=\Omega(E)$ for $0<E<E_c$ 
 because the cut-off does not affects the excitations 
in this energy region.
Furthermore,  $W(E)=0$ for $E > 2 E_c (N_c +1)$ beyond 
an upper bound, where $N_c= E_c/\delta \varepsilon$.  
Equation (\ref{eq:inv_Laplace})  
can be rewritten taking the path along the imaginary axis 
as
\begin{align}
\!
W(E)=
\!\int_{-\infty}^{\infty}  \!\! \frac{dt}{2\pi}
\, e^{i E t} 
\frac{\left(
1+e^{-it E_c}\right)^2
}{4} 
e^{\frac{\scriptstyle 4}{\scriptstyle \delta\varepsilon} 
\frac{\scriptstyle \phi(0)- \phi(it E_c) }{\scriptstyle it} }.
\label{eq:W_integral}
\end{align}
Note that $\mbox{Re}\bigl[ \phi(0) -\phi(ix)\bigr] 
= x^2/4$ for $-\pi <x <\pi$.
% $\phi(ix)$ is a periodic function of real $x$.
Thus, 
 $W(E)$ has a peak at $E =E_c (N_c+1)$,
and around the peak it may be approximated in a Gaussian form  
\begin{align}
\!\!
W(E)
\approx
\frac{4^{2N_c}}{E_c} 
\sqrt{\frac{3}{\pi (2N_c+3)}} 
\  e^{
- \frac{3}{(2N_c+3)} \left(\frac{E}{E_c}-N_c-1\right)^2 
}.
\label{eq:W_integral_app}
\end{align}
This approximate form overestimates $W(E)$ for small $E$, suggesting 
that the real density given by Eq.\ (\ref{eq:W_integral}) has 
larger weight around the peak  than that of the Gaussian.

\end{document}